\documentclass[aip,apl,reprint,graphicx]{revtex4-1}
\usepackage[mediumspace,mediumqspace,squaren]{SIunits}
\usepackage{graphicx}

\begin{document}


\preprint{}

\title{Experimental cross-polarization detection of coupling far-field light to highly confined plasmonic gap modes via nanoantennas} %Title of paper



\author{J. Wen}
\email{Jing.Wen@mpl.mpg.de}
                  \affiliation{Institute of Optics, Information and Photonics, University of Erlangen-Nuremberg}
                  \affiliation{Max Planck Institute for the Science of Light}

\author{P. Banzer}\affiliation{Institute of Optics, Information and Photonics, University of Erlangen-Nuremberg}
                  \affiliation{Max Planck Institute for the Science of Light}

\author{A. Kriesch}\affiliation{Institute of Optics, Information and Photonics, University of Erlangen-Nuremberg}
                   \affiliation{Max Planck Institute for the Science of Light}
\author{D. Ploss}  \affiliation{Institute of Optics, Information and Photonics, University of Erlangen-Nuremberg}
                   \affiliation{Max Planck Institute for the Science of Light}
\author{B. Schmauss}\affiliation{Chair for High Frequency Technology, University of Erlangen-Nuremberg}

\author{U. Peschel} \affiliation{Institute of Optics, Information and Photonics, University of Erlangen-Nuremberg}
                    \affiliation{Max Planck Institute for the Science of Light}

%\thanks{}


\date{\today}

\begin{abstract}
 We experimentally demonstrate the coupling of far-field light to highly confined plasmonic gap modes via connected nanoantennas. The excitation of plasmonic gap modes is shown to depend on the polarization, position and wavelength of the incident beam. Far-field measurements performed in crossed polarization allow for the detection of extremely weak signals re-emitted from gap waveguides and can increase the signal-to-noise ratio dramatically.
\end{abstract}


\maketitle %\maketitle must follow title, authors, abstract and \pacs

\begin{figure}[htb]
\includegraphics[width=0.37\textwidth,bb= 120 557 353 685]{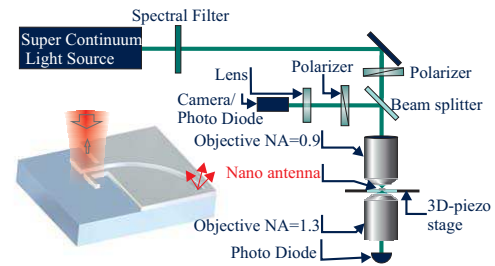}% Here is how to import EPS art
\caption{The configuration of the experimental setup and the schematic view of the measurement.  \label{setup} }
\end{figure}

%\section{}
% The part of preamble
Surface plasmon polaritons (SPPs) are highly confined electromagnetic waves on metal-dielectric interfaces which couple to collective oscillations of the electron gas. Various plasmonic waveguide modes arise from the coupling of SPPs at the interfaces of metals and dielectrics\cite{Maier2005,Verhagen2008,Imre2007,Nomura2005,Chen2006a,Han2010}.
%Various plasmonic waveguide modes were investigated in the past few years.
\citet{Bozhevolnyi2006} reported a channel waveguide mode with a width of \unit{1.1}{\micro\meter} and a propagation length of \unit{100}{\micro\meter}. However, most of the experimentally realized plasmonic modes \cite{Bozhevolnyi2006,Maier2005} have confinements in the \unit{\micro\meter} range and do not reach deeply subwavelength dimensions. Thus they do not have obvious advantages compared to conventional dielectric wave\-guides. In contrast, a true deeply subwavelength gap between two metal surfaces can support an extremely confined mode, but such a compact geometry is more challenging to operate in terms of excitation and detection compared to other structures. \citet{Verhagen2008} reported the near-field detection of metal-insulator-metal (MIM) modes in multilayer structures which showed high confinement perpendicular to the sample surface. Usually, standard phase-matching techniques are employed to excite SPPs, such as Kretschmann, Otto and grating configurations. However, other excitation techniques need to be developed for launching SPPs in deeply subwavelength dimensions.  For the excitation of less-confined stripe wave\-guide modes, generators like arc-shaped nano\-dots and slits were reported by \citet{Imre2007} and \citet{Nomura2005}. Recently, two-dimensional, highly confined gap modes were experimentally realized by \citet{Chen2006a} and \citet{Han2010}. They showed the integration of slot wave\-guides with silicon dielectric waveguides via tapered couplers. However, the aforementioned excitation schemes often lack the compactness that is required for integration into subwavelength, plasmonic circuitry. Nanoantennas are highly compact and were proposed as a kind of nano-coupler to couple far-field light to two-dimensional, highly confined gap modes\cite{Wen2009,Huang2009a}. Recently, the idea of optical wireless interconnects via nanoantennas was  theoretically proposed by \citet{Alu2010}. Thus loading the nanoantennas with plasmonic waveguides is the first step toward optical wireless communication analogous to the ubiquitous loading of radio antennas with coaxial cables in the regime of radio frequency. In this paper, we experimentally demonstrate the excitation of a gap mode enhanced by a nanoantenna and investigate its spatial and spectral dependence on the antenna by cross-polarization detection (see Fig.~\ref{setup}). We prove that optical antennas can potentially be used for selective excitation of highly confined, plasmonic wave\-guide modes in further integrated plasmonic chips.

\begin{figure}[htb]
\includegraphics[width=0.39\textwidth,bb= 133 510 391 672]{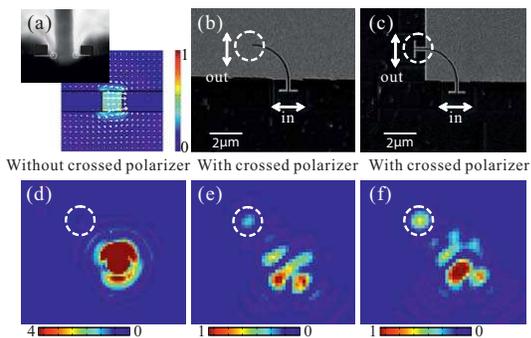}
% Here is how to import EPS art
\caption{(a) SEM image of a cross section of the waveguide and the corresponding mode profile of the electric field (FEM). Optical excitation of gap waveguide modes via nanoantennas and detection of the light backscattered by the waveguide end (b,d,e) or an emitting antenna (c,f). Panels (b,c): SEM images. White arrows indicate the polarization direction of the incident beam and the direction of the crossed polarizer in front of the camera. Panels (d-f): optical reflection camera images with (e,f) and without (d) a crossed polarizer in front of the CCD camera. ($\lambda=\unit{1.5}{\micro\meter}$). The residual spatial intensity pattern of the optical beam in the crossed polarization scheme shown in panels (e,f) was explained by \citet{Mansuripur1986}.
\label{cross_polarization} }
\end{figure}

% The part of the fabrication
For the experiments a \unit{100}{\nano\meter}-thick silver layer with a surface roughness of around \unit{1}{\nano\meter} was deposited on a glass substrate using a magnetron sputtering machine. A chromium layer with a thickness of $1-$\unit{2}{\nano\meter} was used as a wetting layer to improve the quality of the silver layer. Afterwards $80-$\unit{100}{\nano\meter}-wide grooves were structured into the silver layer using a focused ion beam milling system (FIB). A scanning electron microscope (SEM) image of a cross section of the waveguide and the corresponding numerically calculated mode profile (via finite-element method, FEM) of the electric field are given in Fig.~\ref{cross_polarization}(a). Figure~\ref{cross_polarization}(b) shows an SEM image of the measured structure containing a bent-gap wave\-guide with a receiving antenna. Figure~\ref{cross_polarization}(c) shows the same structure but includes an additional transmitting antenna at the other end of the bent wave\-guide. The gap width is \unit{80}{\nano\meter}. The optimized total length of the antenna is designed for a resonance at around $\lambda=\unit{1.5}{\micro\meter}$.

% The part of cross polarization detection
The configuration of the experimental setup is shown in Fig.~\ref{setup}. For excitation, a linearly polarized Gaussian beam (polarization parallel to the antenna) was focused onto the substrate surface. Due to strong back reflection of the incident beam, it is hard to investigate the field in the waveguide unless its direction of polarization is turned so that it becomes visible in the cross-polarized direction. Since the polarization of gap modes is always perpendicular to the tangent of the waveguide, the polarization of the electric fields at the waveguide end can be turned by $90$ degrees via a waveguide bend. Thus, the dominant part of the back reflection of the incident beam was filtered out whereas the scattered light from the very end of the waveguide was detected. For comparison, Fig.~\ref{cross_polarization}(d) shows the reflection image of the sample surface recorded by a camera without a crossed polarizer in front of the camera. In Fig.~\ref{cross_polarization}(e) and (f) two cases with a crossed polarizer in front of the camera are shown. The large bright spot in Fig.~\ref{cross_polarization}(d) is caused by the reflection of the focused beam by the nano antenna. On the upper-left side of the reflected beam, there is only a tiny weak spot with strong background noise. However, for crossed-polarizer detection, a small but clear spot can be observed (see spots inside the white-dashed circles in Fig.~\ref{cross_polarization}(e) and (f)). This demonstrates that the focused optical beam is coupled to the highly confined mode of the plasmonic waveguide via the resonant receiving antenna. Apparently, cross-polarization detection increases the signal-to-noise ratio (SNR) dramatically. In addition, this proves that we have high transmission through bends with \unit{\micro\meter}-sized radii. The emission intensity with a second transmitting antenna (see Fig.~\ref{cross_polarization}(c) and (f)) is enhanced by at least a factor of $2$ compared to the case with just an open end (see Fig.~\ref{cross_polarization}(b) and (e)). The observed scattering at the end of the waveguide weakens and shifts as expected if the sample is moved relative to the optical beam. For an excitation with an incident polarization perpendicular to the long axis of the antenna, at a non-resonant wavelength or without a receiving antenna (reflection images not presented here), no light is observed at the end of the waveguide.
%Consequently, coupling of the light from the far field to the gap plasmonic modes can only be successfully %achieved if the optical beam is on the antenna and at resonant wavelength with the correct polarization direction.

\begin{figure}[htb]
\includegraphics[width=0.34\textwidth,bb=181 526 401 647]{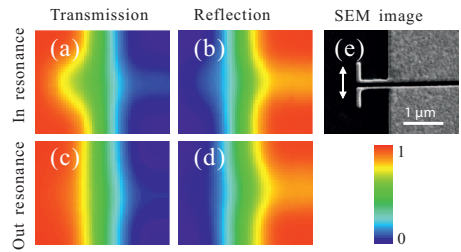}% Here is how to import EPS art
\caption{Scanning measurements in transmission (a) and reflection (b) at resonant wavelength of \unit{1.5}{\micro\meter}. Scanning measurements in transmission (c) and reflection (d) at non-resonant wavelength of \unit{1}{\micro\meter}. (e) SEM image of the corresponding scanning area including the antenna and part of the straight waveguide. Polarization of the incident beam (white arrow) is parallel to the antenna axis. All scans are acquired without a detector polarizer. \label{TRimage} }
\end{figure}

\begin{figure}[htb]
\includegraphics[width=0.38\textwidth,bb= 16 7 198 160]{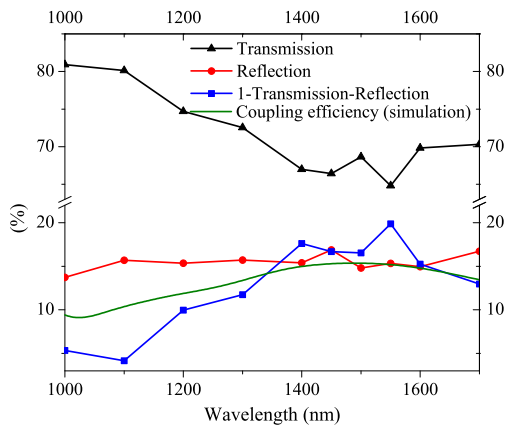}% Here is how to import EPS art
\caption{Experimentally measured spectra in transmission and reflection (normalized). The simulated coupling efficiency of the antenna-waveguide system is shown in the wavelength range between \unit{1}{\micro\meter} and \unit{1.7}{\micro\meter}.  \label{TR} }
\end{figure}

\begin{figure}[htb]
\includegraphics[width=0.38\textwidth,bb= 129 470 339 639]{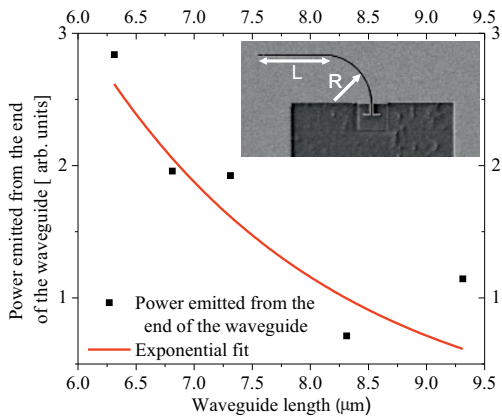}% Here is how to import EPS art
\caption{Power emitted from the end of the waveguide versus total waveguide length, which is fit exponentially (solid line). Inset: SEM image of one measured sample. The radius of the bent part (R) is \unit{3}{\micro\meter}. The length of the straight part (L) varies from \unit{1}{\micro\meter} to \unit{4}{\micro\meter}. \label{decay_loss} }
\end{figure}

%Scanning transmission and reflection images
Figure~\ref{TRimage} shows scans measured in transmission and reflection at the resonant wavelength of \unit{1.5}{\micro\meter} in (a), (b) and at a non-resonant wavelength of \unit{1}{\micro\meter} in (c), (d). For each scan, the incident polarization was parallel to the antenna. No polarizer was situated in front of the detecting photo diode. Each scanning pixel value corresponds to the normalized transmission and reflection measured for a certain position of the focused beam relative to the sample. The sample was scanned through the beam using a defined step size. Transmission and reflection intensities were measured for every position. All presented scans were normalized to the same level. The transmission for resonant excitation of the antenna is lower (yellow color in Fig.~\ref{TRimage}(a)) than the value at the non-resonant wavelength (red color in Fig.~\ref{TRimage}(c)). In the same situation, the reflection does not vary much (see Fig.~\ref{TRimage}(b) and (d)). This already hints that light is coupled into the waveguide for resonant excitation.
% The part of the transmission and reflection spectra, and scanning optical microscopy
We also measured the transmission and reflection spectra of the antenna for incident wavelengths ranging from \unit{1}{\micro\meter} to \unit{1.7}{\micro\meter} (see Fig.~\ref{TR}). The transmission drops by about $10\%$ close to $\lambda=\unit{1.5}{\micro\meter}$, but the reflection stays almost constant for all measured wavelengths. A quite intuitive explanation suggests that, at resonance, energy is coupled to the waveguide mode and results in a reduced transmission but leaves the reflection almost unaffected. For comparison, we simulated the coupling efficiency using the COMSOL Multiphysics commercial FEM package. In order to get a continuous spectrum, the transient analysis module was used. An excitation port which generated an ultrashort electric pulse (spectral width $\unit{1.8}{\micro\meter}$, central wavelength $\unit{1.54}{\micro\meter}$) was set at the end of the waveguide far away from the antenna. The detection port with an area of $\unit{1.5}{\micro\meter} \times \unit{1.5}{\micro\meter}$ ($40 \times 40$ grid) was placed $\unit{200}{\nano\meter}$ above the antenna in air. The powerflow at each point of the detection port in the frequency domain was calculated from the fourier transform of the electric and magnetic fields. The powerflow was then integrated over the detection port area and normalized to the value on the port at the connection between the antenna and the waveguide with an area of $\unit{200}{\nano\meter} \times \unit{200}{\nano\meter}$. This is to exclude the dispersion effect of the waveguide. Suppose the whole process is reciprocal, we thus obtain the maximum coupling efficiency for an excitation from air. The simulated coupling efficiency shows a maximum around \unit{1.5}{\micro\meter}, matching the experimental resonance quite well. Experiments and simulations suggest that about 15\% of the beam's energy is coupled to the waveguide. It should be noted that the calculated coupling efficiency for an excitation from underneath through the substrate is $1.3$ times higher.

% The part of measurement of decay losses
Since we were able to control the waveguide excitation, we also determined the decay length of gap modes with a width of \unit{80}{\nano\meter} at $\lambda=\unit{1.5}{\micro\meter}$. As shown in Fig.~\ref{decay_loss}, the radius of the bent part (R) is fixed to \unit{3}{\micro\meter}, and the length of the straight part (L) varies from \unit{1}{\micro\meter} to \unit{4}{\micro\meter}. The emission from the waveguide end was used for detection. The light emitted by the waveguide end was extracted and normalized to the power of the optical beam focused on the pure glass substrate. The coupling efficiency of the antenna on each measured sample was maximized by adjusting the position of the nanoantenna relative to the focused beam using a piezo stage. By fitting an exponential decay function to the derived data, the decay length was found to be $2.1\pm$\unit{0.8}{\micro\meter}. The rather low value of the decay length might be due to the fact that the fabricated waveguide width of \unit{80}{\nano\meter} is expected to be near cutoff.
%In the experimental samples, the gap was found to be etched down around \unit{30}{\nano\meter} into the glass substrate by cutting the waveguide using the FIB and observing the cross section of the gap. The real cutoff width should be slightly larger than the aforementioned value of \unit{60}{\nano\meter}.
 Also the losses of sputtered silver layers seem to be much higher than the values for silver \cite{Palik1985} used in the simulation. Another reason for the low decay length might be that the FIB structuring process introduces roughness on the sidewalls of the gap, which results in additional scattering losses.

% The conclusion
In summary, the excitation and detection of plasmonic gap modes has been demonstrated. By using a crossed-polarizer detection scheme that requires bent waveguides, the signal-to-noise ratio was dramatically improved. The excitation of gap modes was shown to be sensitive to the wavelength and position of the excitation beam. The measured sum of the coupling efficiency and antenna absorption reached up to 20\% compared to the simulated optimum coupling efficiency of 15\%.
\vspace{1\baselineskip}

%\\[3cm]
The author acknowledges the International Max-Planck Research School (IMPRS) for Optics and Imaging, the Cluster of Excellence for `Engineering of Advanced Materials' (EAM) at the University of Erlangen-Nuremberg and erlangen graduate School in Advanced Optical Technologies (SAOT).

\end{document}